# Size-reduction of nanodiamonds via air oxidation

T. Gaebel[1*], C. Bradac[1], J. Chen[2], P. Hemmer[2], and J.R. Rabeau[1*]

1. ARC Centre of Excellence for Engineered Quantum Systems (EQUS) and Centre for Quantum Science and Technology, Department of Physics and Astronomy, Macquarie University, Sydney, New South Wales 2109, Australia.

2. Electrical and Computer Engineering, Texas A&M University, College Station, TX 77843, USA.

* Corresponding authors: Email: torsten.gaebel@mq.edu.au, james.rabeau@mq.edu.au

**Abstract**

Here we report the size reduction and effects on nitrogen-vacancy centres in nanodiamonds by air oxidation using a combined atomic force and confocal microscope. The average height reduction of individual crystals as measured by atomic force microscopy was 10.6 nm/h at 600 °C air oxidation at atmospheric pressure. The oxidation process modified the surface including removal of non-diamond carbon and organic material which also led to a decrease in background fluorescence. During the course of the nanodiamond size reduction, we observed the annihilation of nitrogen-vacancy centres which provided important insight into the formation of colour centres in small crystals. In these unirradiated samples, the smallest nanodiamond still hosting a stable nitrogen-vacancy centre observed was 8 nm.

## 1. Introduction

The use of colour centres in diamond has attracted considerable interest in a range of research fields including quantum technology and biomedical imaging. Colour centres can serve as single photon sources [1] or bright labels for bio tracking [2]. The nitrogen-vacancy centre (NV) in particular, with its unique optical and spin properties, made possible recent demonstrations such as high resolution magnetometry [3, 4], sub diffraction limit optical microscopy [5, 6], and quantum information technology [7, 8].

Some of these applications only require the unique optical and magnetic properties of NV centres, while other applications also require controlled modification of nanodiamond material properties. For instance, reducing the size of the nanodiamond host, or increasing the colour centre concentration for brighter emission can be crucial for bio-related applications. Size reduction or deaggregation of nanodiamonds can be achieved using a milling process with zirconia beads [9], via plasma etching [10], or thermal treatment [11]. In this work we employed thermal oxidation in air because it is a straight forward approach that enables diamond size reduction and simultaneous modification of the surface chemistry of the nanodiamonds in a way that stabilizes the desired NV charge state [12-15]. Ultimately this increases fluorescence emission efficiency and magnetic sensitivity. Air oxidation is simple and inexpensive and does not require the use of any toxic substance or a specific catalyst, like other oxidation techniques, e.g. plasma etching or acid treatment.

Depending on the temperature of air oxidation, different materials will be removed/etched from the diamond surface including water and physisorbed organic impurities, amorphous carbon and graphitic shells and ultimately the $sp^3$ phase of carbon. Air oxidation is particularly effective at removing graphitic surface layers [16] which has the advantage of greatly suppresses unwanted $sp^2$ fluorescence background. Etch rates for chemical vapour deposited diamond films [17-19] and bulk diamond single crystals [20, 21] showed anisotropy depending on the crystal plane [22]. Our measurements indirectly show the same anisotropy because of the random orientation of the nano-crystal to the substrate.

## 2. Materials and methods

The diamond nanocrystals (Microdiamant, MSY 0-0.1 μm) with a mean size of 50 nm distributed between 0 and 100 nm as measured by an atomic force microscope (AFM), were dispersed on a glass cover slip (Menzer-Glaser). The cover slip was laser scribed with a 5 x 5 grid consisting of 50 x 50 μm² squares (Fig. 1). This enabled the identification of the same nanodiamond sites over consecutive oxidation steps. The sample fluorescence was simultaneously measured with a confocal sample-scanning fluorescence microscope (100× oil immersion objective lens, NA 1.4), excited with a 532 nm CW diode pumped solid-state laser (Coherent, model: Compass 315-M100), and a commercial atomic force microscope (NT-MDT Ntegra) as described in [23, 24] (sketch of the setup see Fig 1a). Intensity autocorrelation curves were measured with a Hanbury Brown and Twiss interferometer setup consisting of 2 avalanche photodiodes (Perkin Elmer) and a correlator (Picoquant).

The oxidation process was carried out in a tube furnace (Lenton thermal designs) in air at atmospheric pressure. The sample consisted of diamond nanocrystals dispersed on a glass coverslip and placed on a metal holder inside the furnace. The furnace temperature was stabilised before inserting the sample into the heated region. The annealing time was measured from the instant the thermocouple, which was mounted in contact with the metal holder, read the same temperature as the internal furnace sensor. All annealing cycles were performed at 600 °C.

# 3. Results and Discussion

The confocal/AFM system and laser scribed grids used to consistently find and compare the physical and optical properties of the same nanodiamonds after each annealing cycle are illustrated in Figure 1(a). After each annealing cycle select squares within the grid were analysed by confocal and AFM measurement. Similar to our previous work [23, 24] this enabled correlation of the nanodiamond fluorescence and crystal topology.

The nanodiamonds were all treated with a preliminary 2 h heating step at 600 °C to remove the relatively large amount of non-diamond carbon, as well as other impurities on the sample. The fact that the surface tended to have a high proportion of $sp^2$ was confirmed in the etch rate measurements: where an increased etch rate was measured in the early stage of annealing (see Fig. 3b). This can be explained by the higher etch rate of $sp^2$ compared to $sp^3$ carbon [16]. Figure 2 shows the spectra for a NV centre taken before and after annealing. Before the treatment a broad unstructured fluorescence ranging from 550 nm to 800 nm was observed, which we attribute to a high graphite content at the surface of the nanodiamond (spectra is similar to [25], which is attributed to surface defects and graphite). After 1h of heating, the spectral features originating from NV are much more pronounced. Subsequent annealing cycles do not change the spectra in a noticeable way.

The autocorrelation curve shows in a similar way an increase in the visibility of the antibunching dip, which means that the background fluorescence level decreased after the first air oxidation step. This also supports our assertion of the preferential removal of graphitic and other non-diamond carbon material in the early oxidation stage [16], after which the etching is almost exclusively of diamond.

A histogram of the height of individual crystals was acquired from the AFM measurements after consecutive annealing steps which gave an indication of the average change in size as a function of anneal time. The height values were obtained by fitting 2-dimensional Gaussians to individual crystals on the AFM image after subtracting the background height offset.

By fitting Gaussians to the height distributions we were able to infer a "mean" crystal size after each step. In Fig. 3b the mean sizes are plotted as a function of time and fitted with a line to give the average etch rate. Note that the first step from 0 to 2 h shows a dramatic reduction in crystal size. As discussed, this is due to the rapid etching of non-diamond carbon on the surface and the point is excluded from the linear fit made from 2 h onward at which point the etch rate is constant. The fit gives an etch rate of 10.6 nm/h and was determined from 7 consecutive oxidation steps, each one lasting 30 min at 600 °C. Figure 3a shows the resulting histograms of nanodiamond sizes for 3 different steps within the heating cycles.

When examining the size reduction behaviour of *individual* crystals, one can see that the etch rate varies somewhat. This effect is visible in Fig. 4 where 3 nanodiamonds were tracked over three air oxidation steps. Two of the crystals shrink consistently in height, but one stays quite constant around 50 nm. However, the relatively crude measure of the X and Y cross sections (i.e. the width of the crystal) of this particular nanodiamond as a function of time does show a reduction. The likely reason for this is an anisotropy in the etch rate for different crystallographic planes/surfaces as already observed a in CVD diamond samples [22] and natural diamond [21].

More interestingly with our experimental apparatus we were able to study the size of individual nanodiamond particles hosting NV defects. To do this we required a confocal image of the nanodiamonds and a corresponding AFM image of the same region. In Fig. 4 these images were put together for the untreated sample and two oxidation steps. Not surprisingly one can observe the annihilation of NV defects via air oxidation as layers of carbon are removed eventually exposing and destroying the NV defects themselves.

With these results we can determine the size distribution of the nanodiamonds hosting NV centres (see Fig. 6). Even though the size distribution of the powder used is specified to be ranging from 0 to 100 nm we were able to find particles up to 170 nm in height. This may have been due to aggregation of the nanodiamonds. The air oxidation and the resulting shrinking of the nanodiamonds helped to minimise aggregation, as shown in Fig. 3. The smallest nanodiamond that we observed in this study which still hosted an optically active NV centre was 8 nm in height. To increase the probability of ending up with even smaller nanodiamonds containing NVs, one would have to increase the starting concentration of available NVs, which can be done by implanting the diamond powder with electrons or ions [26]. Notice that in our study the sample was analysed "as received" without undergoing any NV-concentration enhancing process.

**Conclusion**

We have demonstrated an easy to use technique for studying the effect of air oxidation on the size of nanodiamonds and luminescent properties of contained NV centres. The smallest nanodiamond still hosting an NV centre we found was 8nm. Determining the NV size limits and surface effects is a crucial step to understand and eventually be able to control the incorporation of NV centres into small diamonds.

**Acknowledgement**

We acknowledge support from the Australian Research Council Future Fellowship scheme (FT0991243). CB is supported by an MQ Research Excellence International Scholarships and TG is funded by a MQ Research Fellowship. PH and JC acknowledge support from NIH #C09-00053. Substrate grids were fabricated by the ANFF funded Optofub at Macquarie University (B. Johnston).

# Figure Captions

**Figure 1.** Experimental characterization setup. (a) shows an artistic view of the confocal beam incident from the bottom and through the glass coverslip combined with the AFM tip probing the sample from above. The inset is a photograph of the sample from directly above, one can see the laser scribed grid and the AFM cantilever. (b) is a confocal fluorescence intensity map of the sample and (c) shows the corresponding AFM Height map. The scan area is 50 x 50 μm.

**Figure 2.** (a) normalised spectra of an NV centre following a 1h oxidation step. The black line is the spectrum without air oxidation, and the green one shows the spectrum after 1h air oxidation. (b) shows the corresponding $g^2$ curves. Note that the contrast of the antibunching feature increases, because of a decrease in background luminescence.

**Figure 3.** Size reduction as a function of time of nanodiamonds treated in air at 600°C. (a) Histograms of the nanodiamond sizes after a specific oxidation time. These distributions are measured consecutively on the same sample. The solid lines are Gaussian fits to the size distribution. (b) The mean of the size distributions plotted over time. A linear fit (solid line) indicates an etch rate of 10.6 nm/h.

**Figure 4.** 3-dimensional AFM images of the same nanodiamonds following the oxidation steps. The crystal at the centre did not experience a reduction in height as much as the other 2 crystals.

**Figure 5.** Confocal and AFM images taken after consecutive oxidation steps. (a) confocal image taken before oxidation; the numbered circles indicate some of the NV centres we selected for the analysis. (b) confocal image after annealing the sample in the furnace in air for 2.5 hours at 600 °C; the red-numbered circles indicate NV centres which annihilated. (c) confocal image after annealing the sample in the furnace for another 2.5 hours at 600 °C. The *-numbered squares indicate NV centres which were created by the heating process due to vacancy diffusion to existing $N_s$ sites. (d) AFM image taken before oxidation; the circles indicate some of the crystal we selected for the analysis. (e) AFM image after annealing the sample in the furnace for 2.5 hours at 600 °C. (f) AFM image after annealing the sample in the furnace for another 2.5 hours at 600 °C. The insets show the reduction in size for the highlighted crystal. The square box shows an example of diamond crystals going from being clustered together to being isolated.

**Figure 6.** Size distribution of the nanodiamonds hosting NV centres.

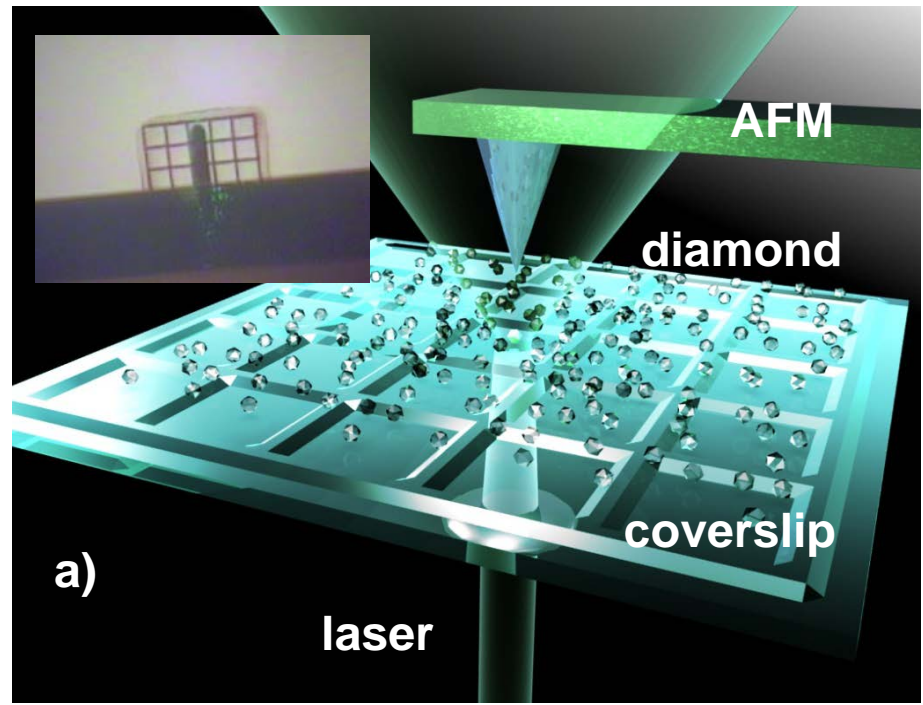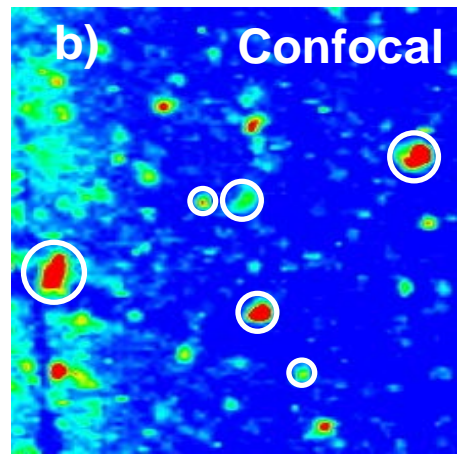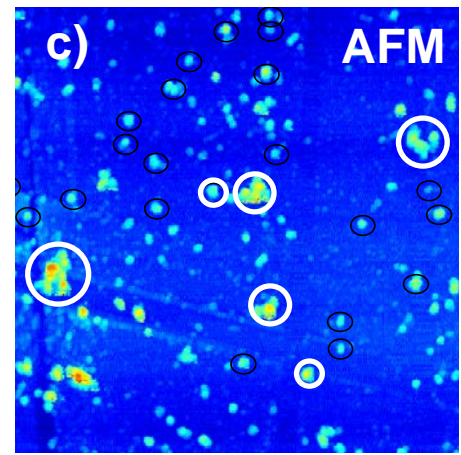

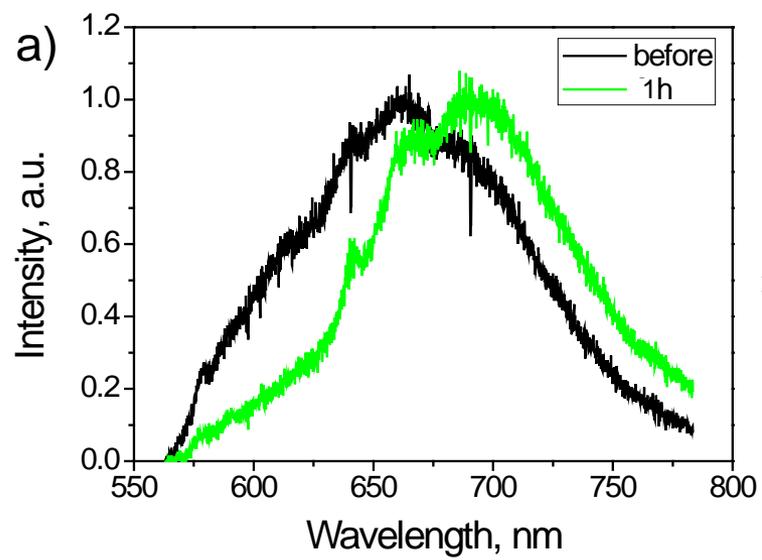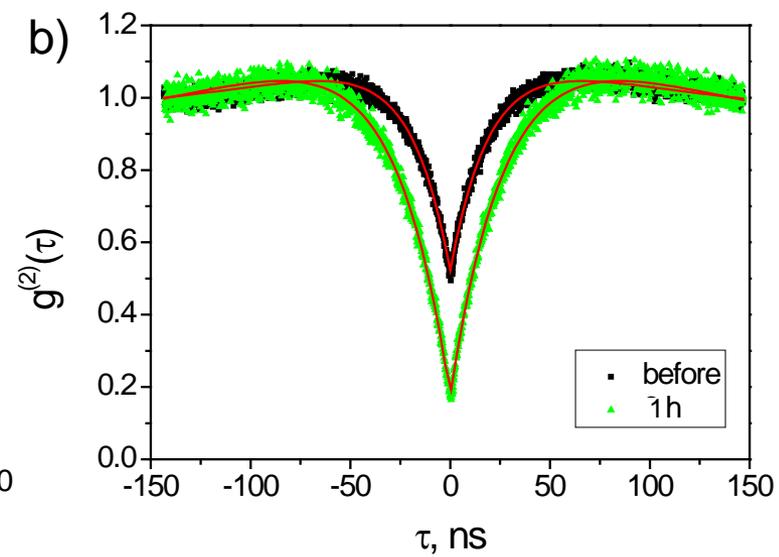

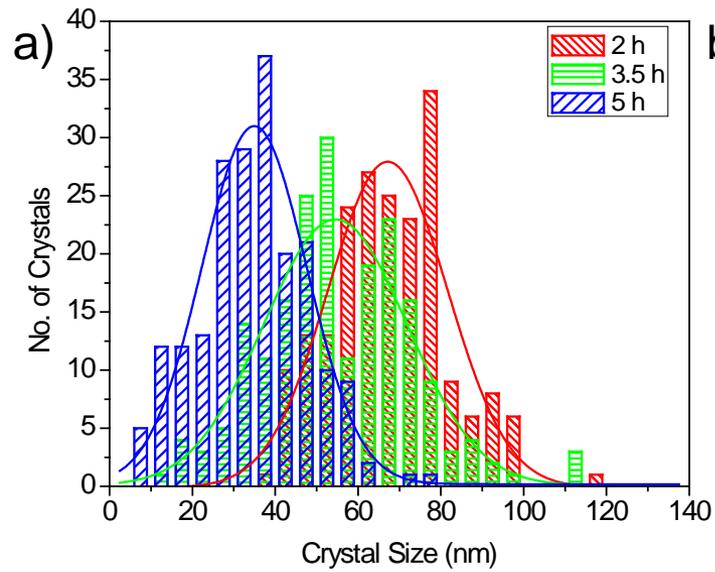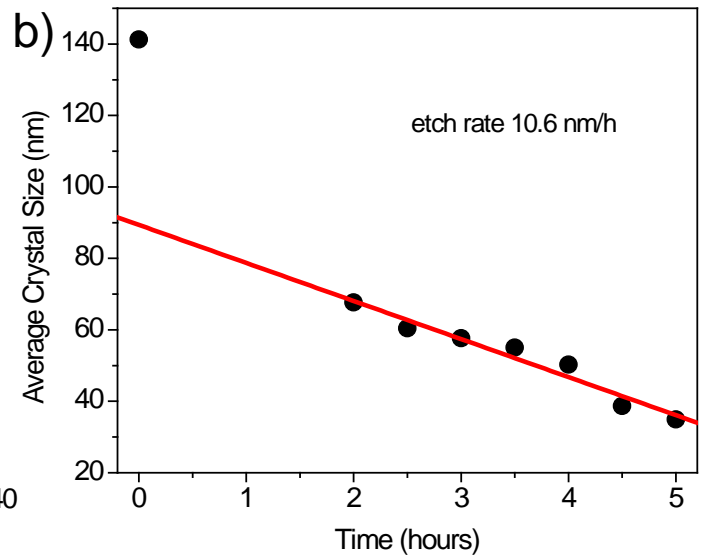

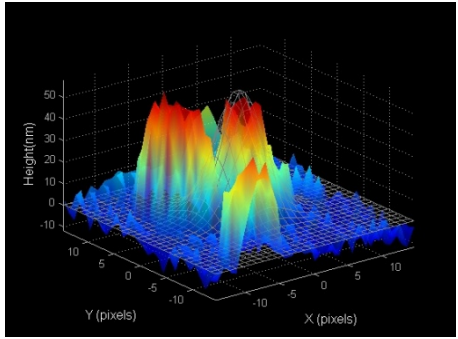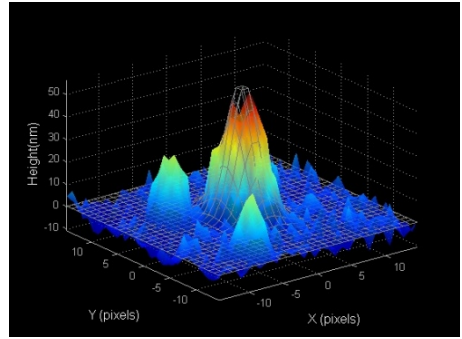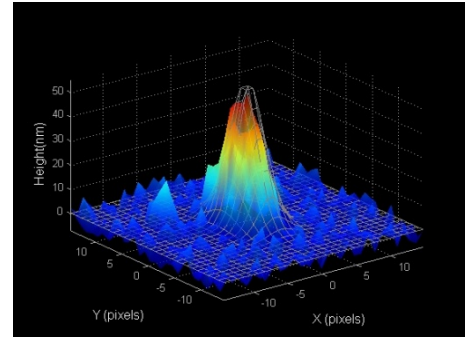

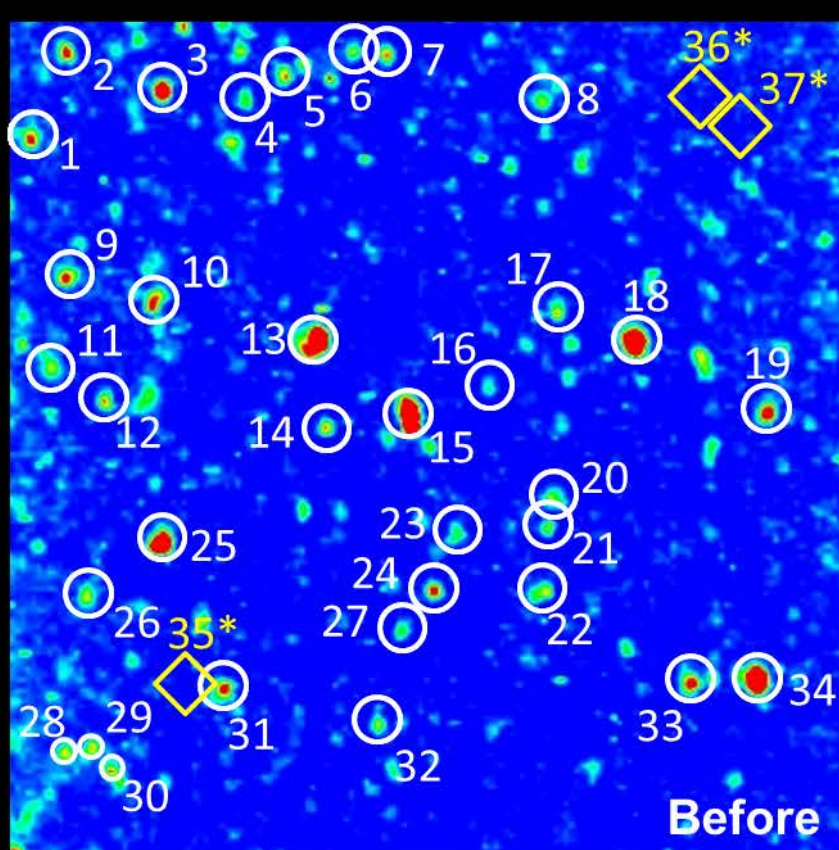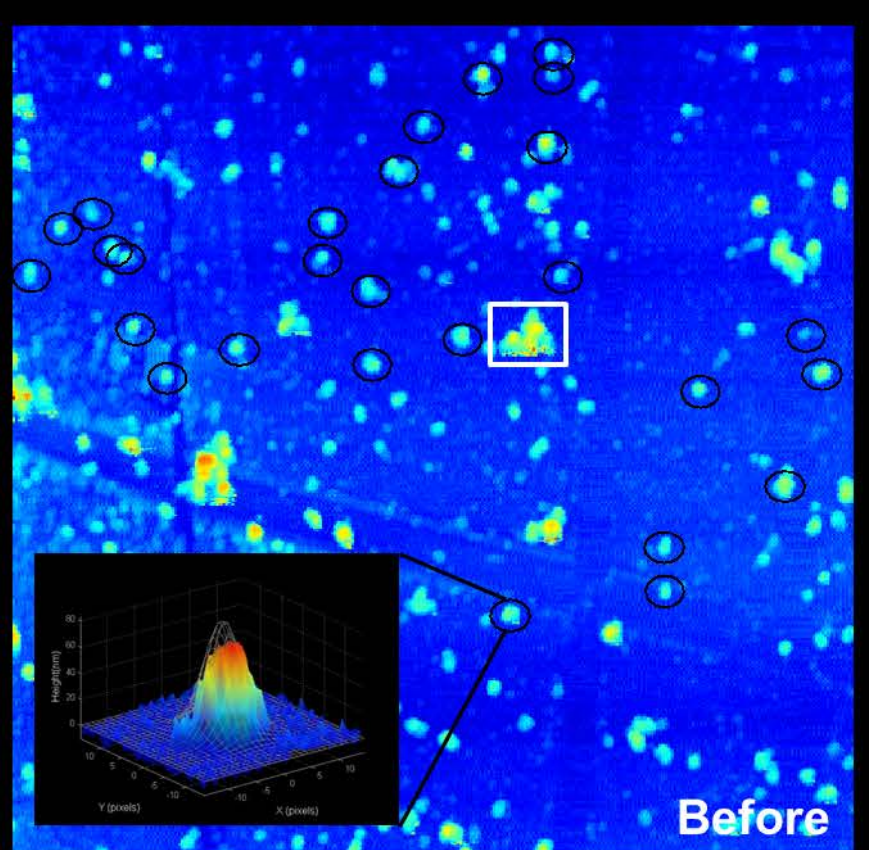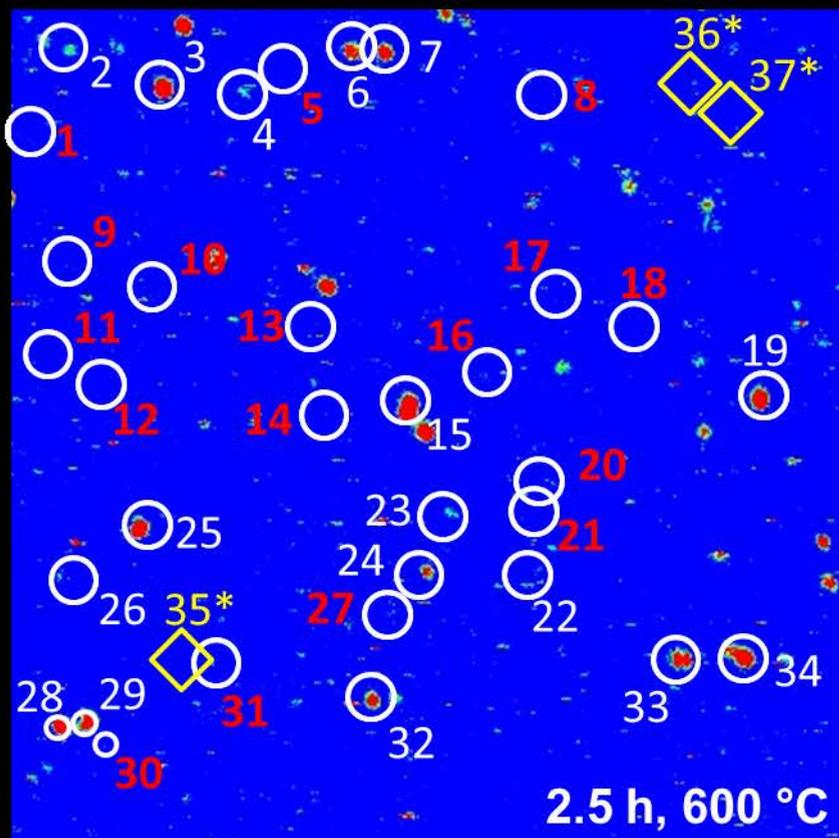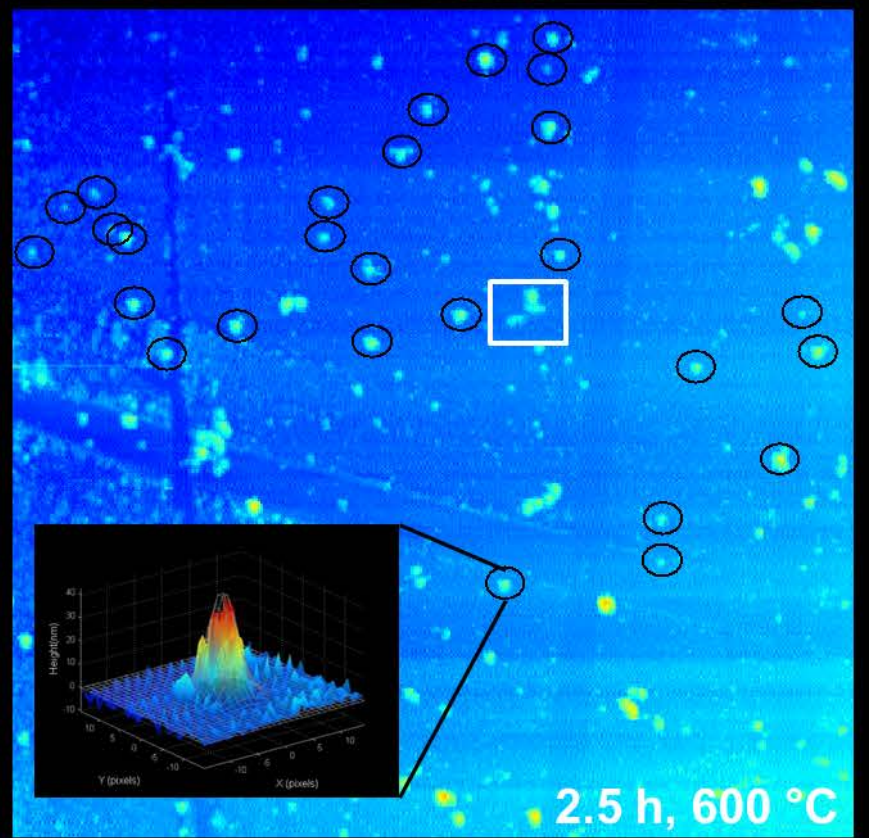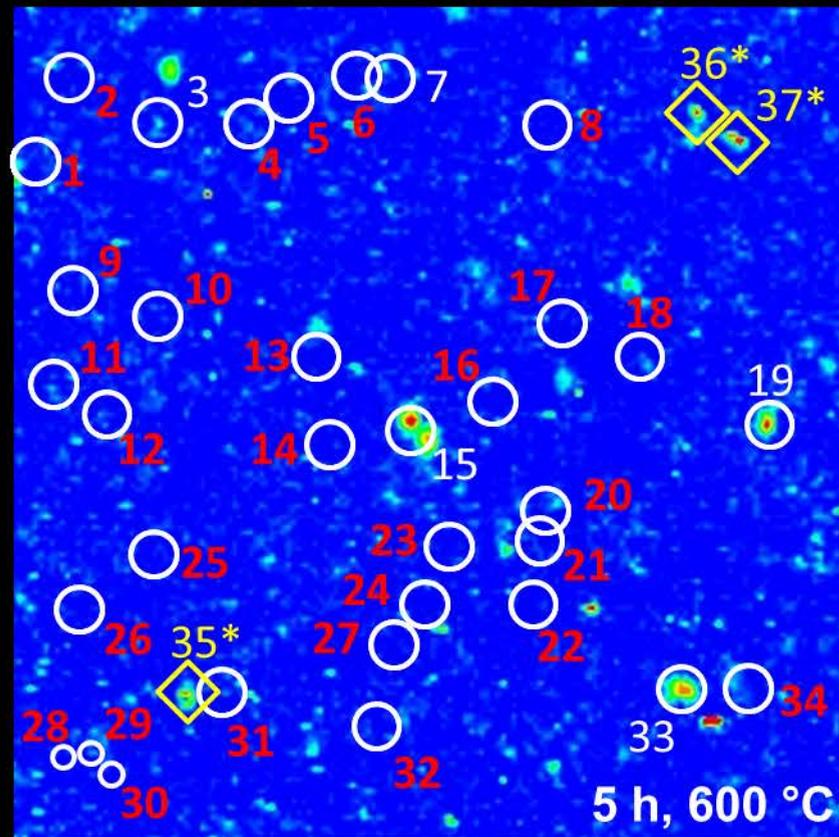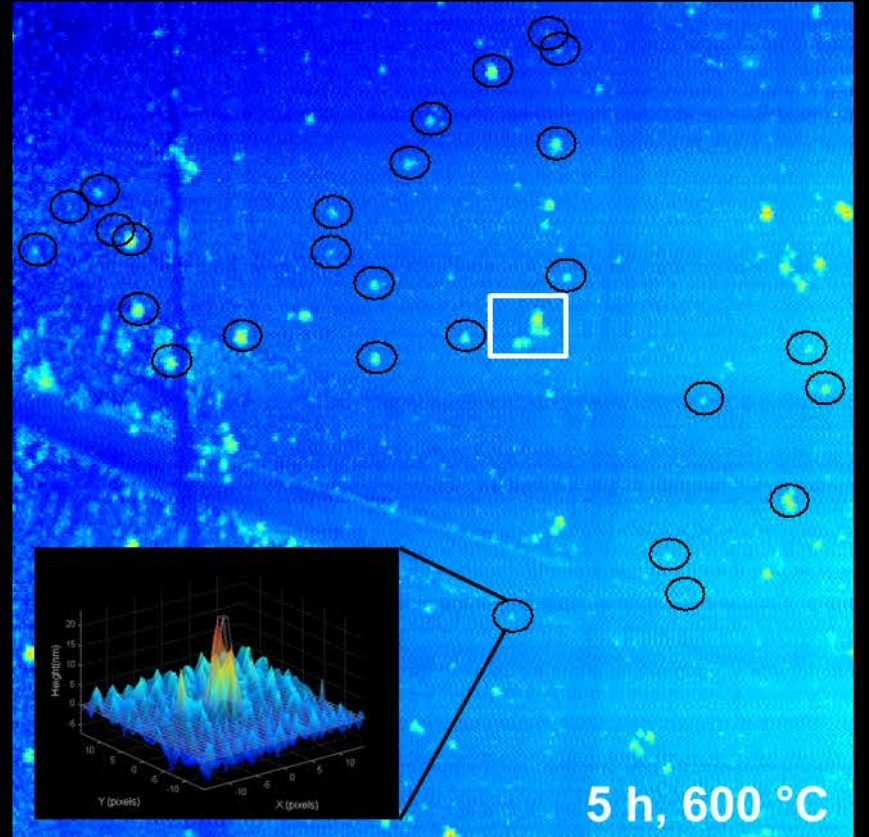

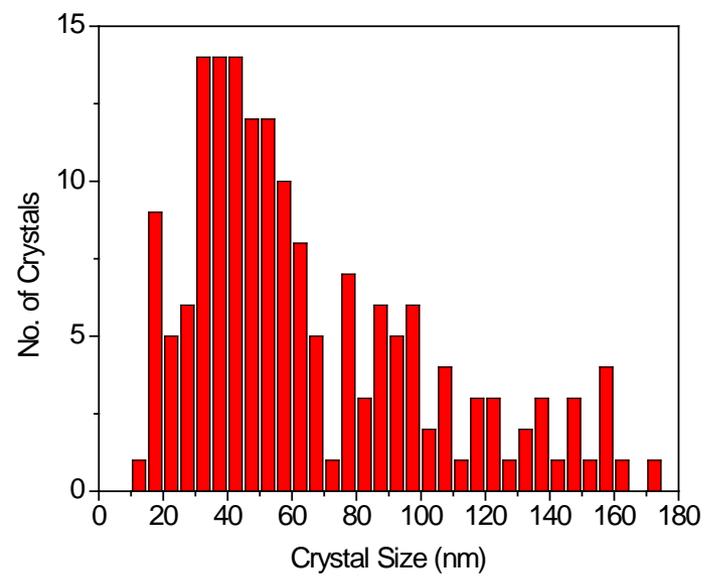